\documentclass[fleqn,10pt]{wlscirep}
\usepackage[utf8]{inputenc}
\usepackage[T1]{fontenc}
\usepackage{gensymb}
\usepackage{tabularx}
\usepackage{multicol,array}
\usepackage{subcaption}
\usepackage{tikz}
\usepackage[font=small,justification=justified,format=plain]{caption}

\newcolumntype{C}[1]{>{\centering\arraybackslash}p{#1}}

\title{High-Index Semiconductor Nanoparticles as Low-Loss Alternatives to Gold for Refractive Index Sensing}

\author[1]{Bernat Frangi}

\begin{abstract}
This study presents a comparative numerical analysis of Gold (Au) and high-index semiconductor nanoparticles for refractive index sensing in the visible range. While Au nanoparticles demonstrate high sensitivity ($\approx 150$ nm per refractive index unit), their performance is constrained by ohmic losses. In contrast, high-index dielectrics are shown to exhibit comparable extinction efficiencies driven exclusively by scattering, thereby minimizing thermal losses. Multipolar decomposition reveals that semiconductors support simultaneous electric and magnetic Mie resonances, the interference of which enables directional scattering phenomena unattainable in small metallic particles. These findings suggest that high-index nanostructures offer a robust, low-loss alternative to plasmonics for advanced sensing applications.\\

\end{abstract}

\begin{document}

\flushbottom
\maketitle
\thispagestyle{empty}

\begin{multicols}{2}

\section*{Introduction}

The excitation of Localized Surface Plasmon Resonances (LSPRs) in metallic nanoparticles manifests as a pronounced resonance in scattering and absorption efficiencies. Driven by the collective oscillation of conduction electrons, this phenomenon yields extreme near-field enhancement, subwavelength confinement, and broadband tunability \cite{willets2007localized, doi:10.1021/cr100313v}. The spectral position of the resonance—defined as the wavelength of peak extinction efficiency—depends critically on particle geometry and the refractive index of the surrounding medium, underscoring the potential of these nanoparticles as highly sensitive plasmonic refractive index sensors.

However, a significant limitation of plasmonics in many applications is the presence of substantial ohmic (thermal) losses arising from high absorption \cite{Boriskina:17, Khurgin2015}. Consequently, resonant high-index semiconductor nanoparticles have emerged as a compelling alternative. Unlike metals, which rely on free-electron plasma oscillations and support primarily electric dipolar modes, these dielectric nanostructures exhibit negligible absorption and support both electric and magnetic Mie resonances driven by displacement currents \cite{Kivshar:17, kuznetsov2016optically}. This substitution of ``plasmonic'' resonances with low-loss dielectric Mie resonances offers comparable tunability based on particle size and environmental refractive index, but with the added advantage of exhibiting lower losses and supporting magnetic modes.

In this work, we first investigate the viability of Gold (Au) nanoparticles for refractive index sensing by examining how their resonance characteristics evolve in the visible range for particle radii between 10 and 50 nm within surrounding media ranging from $n=1$ to $n=2$. Gold is selected for this analysis due to its excellent biocompatibility, chemical stability, and established utility in biosensing and biomedical diagnostics \cite{C1CS15237H, saha2012gold}. Subsequently, we extend our study to the resonance features of seven high-index semiconductor nanoparticles with a particle size of 100 nm to compare their characteristics with those of their metallic counterparts.

\section*{Results and discussion}

\subsection*{Au nanoparticles}

\begin{figure*}[htbp]
    \centering
    \begin{subfigure}[t]{0.3\textwidth}
    \begin{tikzpicture}[inner sep=0]
      \node[anchor=south west] (img) at (0,0)
        {\includegraphics[width=\textwidth]{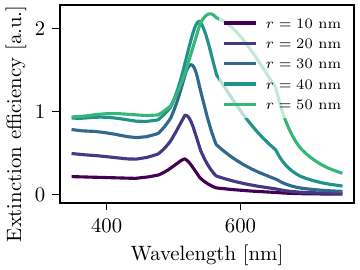}};
      \node[anchor=north west, xshift=-14pt, yshift=-3pt]
        at (img.north west) {\textbf{(a)}};
    \end{tikzpicture}
    \end{subfigure}
    \hfill
    \begin{subfigure}[t]{0.3\textwidth}
    \begin{tikzpicture}[inner sep=0]
      \node[anchor=south west] (img) at (0,0)
        {\includegraphics[width=\textwidth]{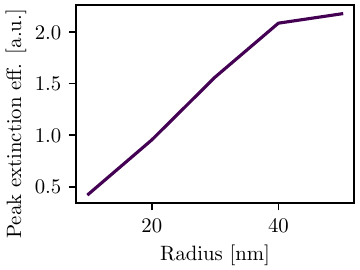}};
      \node[anchor=north west, xshift=-14pt, yshift=-3pt]
        at (img.north west) {\textbf{(b)}};
    \end{tikzpicture}
    \end{subfigure}
    \hfill
    \begin{subfigure}[t]{0.3\textwidth}
    \begin{tikzpicture}[inner sep=0]
      \node[anchor=south west] (img) at (0,0)
        {\includegraphics[width=\textwidth]{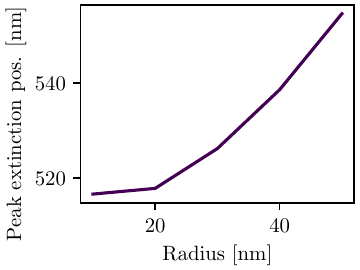}};
      \node[anchor=north west, xshift=-14pt, yshift=-3pt]
        at (img.north west) {\textbf{(c)}};
    \end{tikzpicture}
    \end{subfigure}
    \hfill\hfill
    
    \begin{subfigure}[t]{0.3\textwidth}
    \begin{tikzpicture}[inner sep=0]
      \node[anchor=south west] (img) at (0,0)
        {\includegraphics[width=\textwidth]{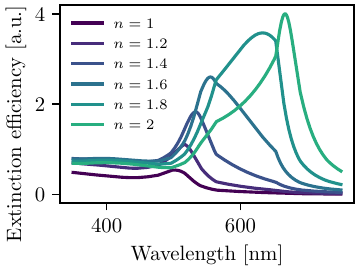}};
      \node[anchor=north west, xshift=-14pt, yshift=-3pt]
        at (img.north west) {\textbf{(d)}};
    \end{tikzpicture}
    \end{subfigure}
    \hfill
    \begin{subfigure}[t]{0.3\textwidth}
    \begin{tikzpicture}[inner sep=0]
      \node[anchor=south west] (img) at (0,0)
        {\includegraphics[width=\textwidth]{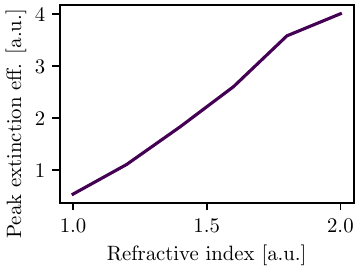}};
      \node[anchor=north west, xshift=-14pt, yshift=-3pt]
        at (img.north west) {\textbf{(e)}};
    \end{tikzpicture}
    \end{subfigure}
    \hfill
    \begin{subfigure}[t]{0.3\textwidth}
    \begin{tikzpicture}[inner sep=0]
      \node[anchor=south west] (img) at (0,0)
        {\includegraphics[width=\textwidth]{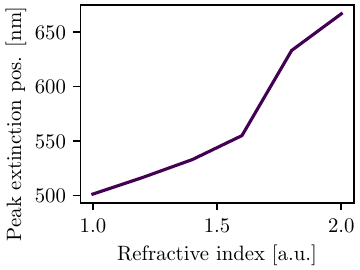}};
      \node[anchor=north west, xshift=-14pt, yshift=-3pt]
        at (img.north west) {\textbf{(f)}};
    \end{tikzpicture}
    \end{subfigure}
    \hfill\hfill
    \caption{(a) Extinction efficiency $Q_\text{ext}$ within the visible range (350 to 750\,nm) for Au nanoparticles with different values of the radius $r$ between 10 and 50\,nm surrounded by a medium with $n=1.333$ (such as water). (b) Peak $Q_\text{ext}$ as a function of $r$, obtained from the cases shown in (a). (c) Spectral position of the peak $Q_\text{ext}$ as a function of $r$, obtained from the cases shown in (a). (d) $Q_\text{ext}$ within the visible range for an Au nanoparticle of 30\,nm radius surrounded by media with different values of $n$ between $n=1$ and $n=2$. (e) Peak $Q_\text{ext}$ as a function of $n$, obtained from the cases shown in (d). (f) $\lambda_\text{max}$ as a function of $n$, obtained from the cases shown in (d).}
    \label{fig:part-1-dep}
\end{figure*}
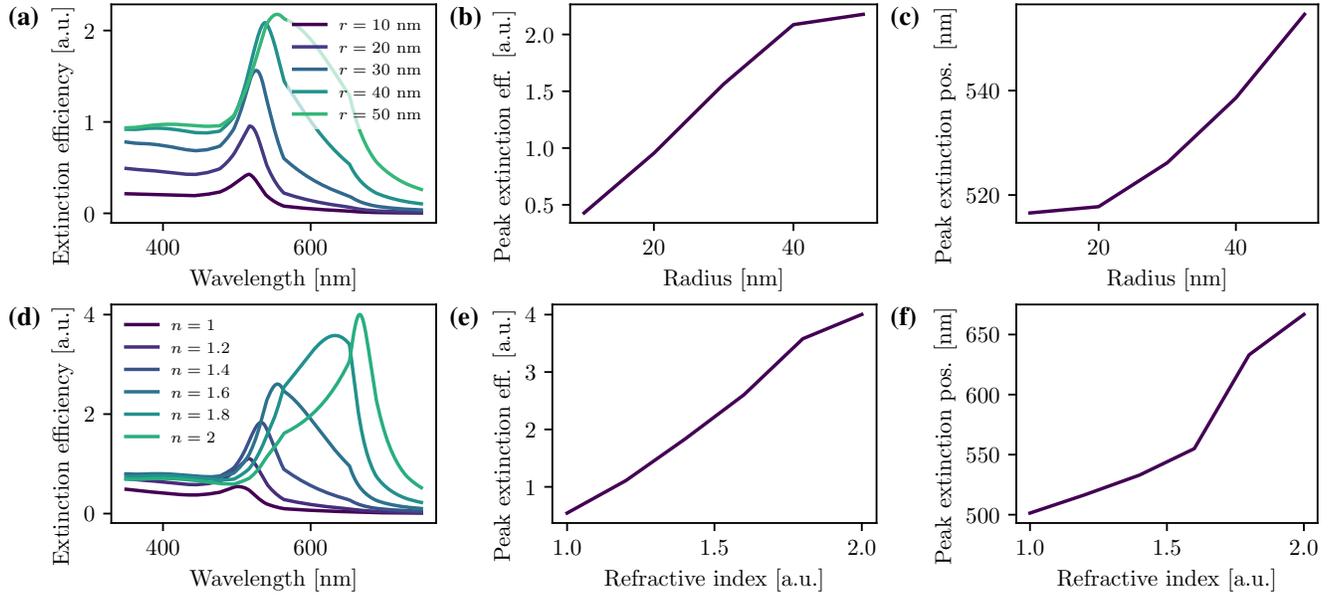

Simulations reveal \cite{bernat_frangi_2025_18069231} that for a Gold nanoparticle in water, the resonant extinction intensifies, red-shifts, and broadens as the particle radius $r$ increases (\textbf{Figure \ref{fig:part-1-dep}a}). Analysis of the scattering coefficients confirms that these features arise exclusively from the electric dipolar mode, as all Mie coefficients other than $a_1$ are negligible. The peak extinction efficiency scales linearly with radii between 10\,nm and 40\,nm, rising from $0.5$ to $2$ before approaching saturation near 50\,nm (\textbf{Figure \ref{fig:part-1-dep}b}). Conversely, the spectral position of the peak follows an exponential dependence on $r$, yielding a total wavelength shift of approximately 40\,nm between $r=10$\,nm and $r=50$\,nm (see \textbf{Figure \ref{fig:part-1-dep}c}).

Analogous simulations indicate that increasing the external medium refractive index $n$ causes the resonant extinction to increase, red-shift, and undergo substantial spectral shape modification (\textbf{Figure \ref{fig:part-1-dep}d}); these features are also driven exclusively by the electric dipolar mode. The peak extinction efficiency exhibits a linear relationship with $n$ between 1 and 2, increasing from $0.5$ to $4$ and approaching saturation near $n=2$ (\textbf{Figure \ref{fig:part-1-dep}e}). The spectral position $\lambda_\text{max}$ of the peak, however, displays a non-trivial dependence on $n$, resulting in a total wavelength shift exceeding 150\,nm between $n=1$ and $n=2$ (see \textbf{Figure \ref{fig:part-1-dep}f}).

Because geometric variations induce spectral changes that can mimic those arising from refractive index fluctuations, distinguishing between size polydispersity and genuine environmental effects remains challenging. Consequently, the effective deployment of these nanoparticles for refractive index sensing requires stringent control over particle size distribution.

With these considerations in mind, a refractive index sensor can be implemented through two distinct modalities. The first approach tracks the peak extinction efficiency $Q_\text{ext}$, yielding an approximate sensitivity of $\Delta Q_{\text{ext}} / \Delta n \approx 3.5$ (estimated from the change of $0.5$ to $4$ over a unit refractive index interval). Alternatively, monitoring the spectral position $\lambda_\text{max}$ suggests a sensitivity of $\Delta \lambda_\text{max} / \Delta n \approx 150$\,nm per refractive index unit (RIU). However, this latter estimate relies on a linear approximation that obscures the observed non-trivial dependence on $n$. To improve measurement robustness, a dual-parameter strategy that combines both intensity and spectral-shift information can be employed to mitigate these non-linearities and yield a more accurate determination of the true refractive index.

\subsection*{High-index semiconductor nanoparticles}

\begin{figure*}[htbp]
    \centering
    \begin{subfigure}[t]{0.3\textwidth}
    \begin{tikzpicture}[inner sep=0]
      \node[anchor=south west] (img) at (0,0)
        {\includegraphics[width=\textwidth]{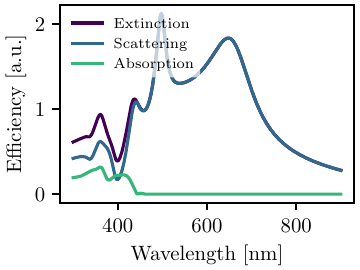}};
      \node[anchor=north west, xshift=-14pt, yshift=-3pt]
        at (img.north west) {\textbf{(a)}};
    \end{tikzpicture}
    \end{subfigure}
    \hfill
    \begin{subfigure}[t]{0.3\textwidth}
    \begin{tikzpicture}[inner sep=0]
      \node[anchor=south west] (img) at (0,0)
        {\includegraphics[width=\textwidth]{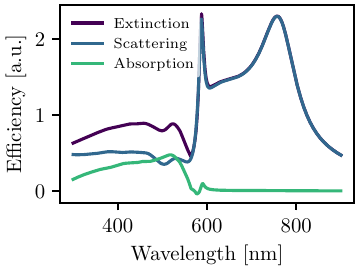}};
      \node[anchor=north west, xshift=-14pt, yshift=-3pt]
        at (img.north west) {\textbf{(b)}};
    \end{tikzpicture}
    \end{subfigure}
    \hfill
    \begin{subfigure}[t]{0.3\textwidth}
    \begin{tikzpicture}[inner sep=0]
      \node[anchor=south west] (img) at (0,0)
        {\includegraphics[width=\textwidth]{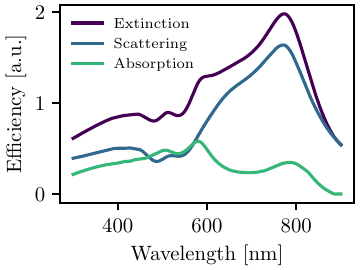}};
      \node[anchor=north west, xshift=-14pt, yshift=-3pt]
        at (img.north west) {\textbf{(c)}};
    \end{tikzpicture}
    \end{subfigure}
    \hfill\hfill
    
    \begin{subfigure}[t]{0.3\textwidth}
    \begin{tikzpicture}[inner sep=0]
      \node[anchor=south west] (img) at (0,0)
        {\includegraphics[width=\textwidth]{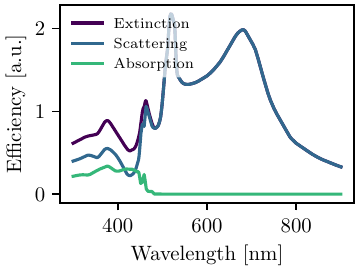}};
      \node[anchor=north west, xshift=-14pt, yshift=-3pt]
        at (img.north west) {\textbf{(d)}};
    \end{tikzpicture}
    \end{subfigure}
    \hfill
    \begin{subfigure}[t]{0.3\textwidth}
    \begin{tikzpicture}[inner sep=0]
      \node[anchor=south west] (img) at (0,0)
        {\includegraphics[width=\textwidth]{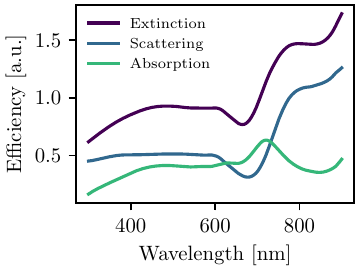}};
      \node[anchor=north west, xshift=-14pt, yshift=-3pt]
        at (img.north west) {\textbf{(e)}};
    \end{tikzpicture}
    \end{subfigure}
    \hfill
    \begin{subfigure}[t]{0.3\textwidth}
    \begin{tikzpicture}[inner sep=0]
      \node[anchor=south west] (img) at (0,0)
        {\includegraphics[width=\textwidth]{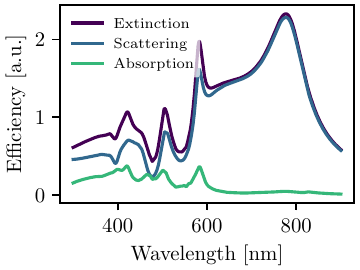}};
      \node[anchor=north west, xshift=-14pt, yshift=-3pt]
        at (img.north west) {\textbf{(f)}};
    \end{tikzpicture}
    \end{subfigure}
    \hfill\hfill
    
    \begin{subfigure}[t]{0.3\textwidth}
    \begin{tikzpicture}[inner sep=0]
      \node[anchor=south west] (img) at (0,0)
        {\includegraphics[width=\textwidth]{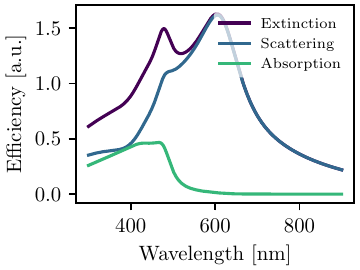}};
      \node[anchor=north west, xshift=-14pt, yshift=-3pt]
        at (img.north west) {\textbf{(g)}};
    \end{tikzpicture}
    \end{subfigure}
    \hfill
    \begin{subfigure}[t]{0.3\textwidth}
    \begin{tikzpicture}[inner sep=0]
      \node[anchor=south west] (img) at (0,0)
        {\includegraphics[width=\textwidth]{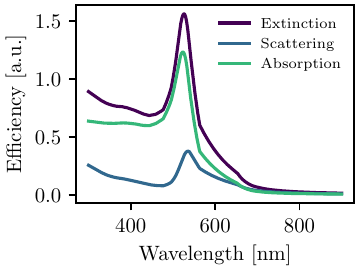}};
      \node[anchor=north west, xshift=-14pt, yshift=-3pt]
        at (img.north west) {\textbf{(h)}};
    \end{tikzpicture}
    \end{subfigure}
    \hfill
    \begin{subfigure}[t]{0.3\textwidth}
    \begin{tikzpicture}[inner sep=0]
      \node[anchor=south west] (img) at (0,0)
        {\includegraphics[width=\textwidth]{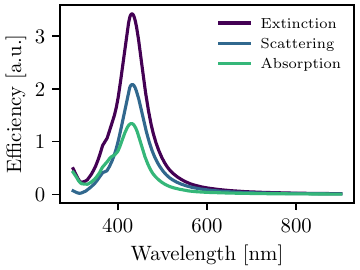}};
      \node[anchor=north west, xshift=-14pt, yshift=-3pt]
        at (img.north west) {\textbf{(i)}};
    \end{tikzpicture}
    \end{subfigure}
    \hfill\hfill
    \caption{Extinction, scattering and absorption efficiencies within the visible range (300 to 900\,nm) for high-index semiconductor nanoparticles of radius 100\,nm made of (a) AlAs, (b) AlSb, (c) GaAs, (d) GaP, (e) Ge, (f) Si and (g) TiO$_2$, and for metal nanoparticles of radius 30\,nm made of (h) Au and (i) Ag. In all cases, the refractive index of the surrounding medium is $=1.333$ (such as water). For high-index semiconductors, it is easy to see that the largest contribution to extinction near the resonances comes from scattering, whereas in metals absorption contributes significantly.}
    \label{fig:part-2-extiction}
\end{figure*}
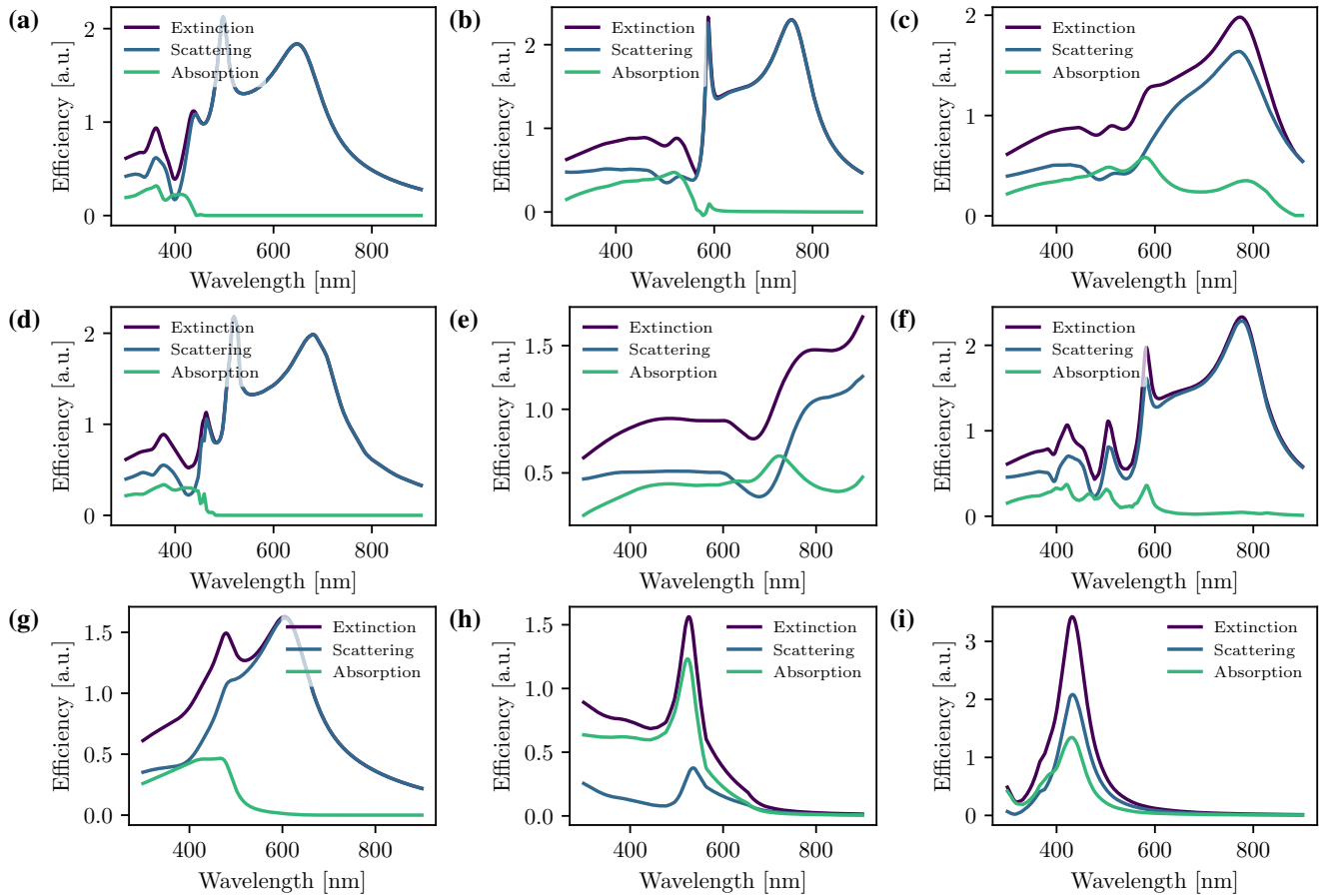

Simulations of seven high-index semiconductors \cite{bernat_frangi_2025_18069231} confirm that dielectric Mie resonances achieve extinction efficiencies comparable to those of Au nanoparticles, despite the absence of plasma oscillations. In contrast to the metallic case, however, the dielectric response is dominated by scattering, thereby exhibiting minimal ohmic losses (see \textbf{Figure \ref{fig:part-2-extiction}}).

\begin{figure*}
    \centering
    \includegraphics[width=0.65\textwidth]{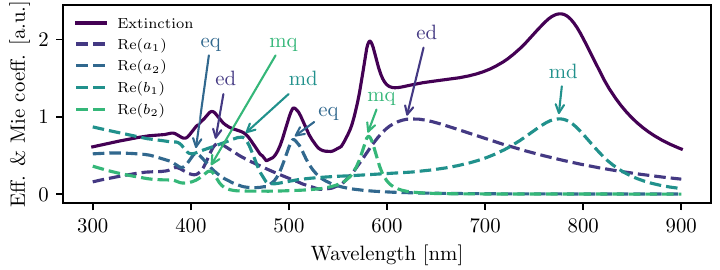}
    \caption{Extinction efficiency and real part of the Mie coefficients $a_1$ (electric dipole), $a_2$ (electric quadrupole), $b_1$ (magnetic dipole) and $b_2$ (magnetic quadrupole) for a Si nanoparticle of 100\,nm radius surrounded by a medium with $n=1.333$ (such as water) within the visible range (300 to 900\,nm).}
    \label{fig:si-extinction}
\end{figure*}

\textbf{Figure \ref{fig:part-2-extiction}} also illustrates that high-index semiconductors exhibit significantly greater spectral complexity than their metallic counterparts. This intricate behavior stems not only from the electric dipole modes observed in metal nanoparticles but also from magnetic dipole modes---generated by circulating displacement currents---and higher-order multipoles (specifically in our analysis, quadrupoles) enabled by the larger particle dimensions. To elucidate the origin of these resonant features, \textbf{Figure \ref{fig:si-extinction}} presents a decomposition of the extinction efficiency for a Si nanoparticle. The analysis confirms the coexistence of electric and magnetic modes, which manifest as both dipole and quadrupole resonances due to the increased size of the nanoparticle.

\begin{figure*}[ht!]
    \centering
    \hfill
    \begin{minipage}[t]{0.32\textwidth}
        \vspace{0pt} 
        \begin{tikzpicture}[inner sep=0]
            \node[anchor=south west] (img) at (0,0) {\includegraphics[width=\textwidth]{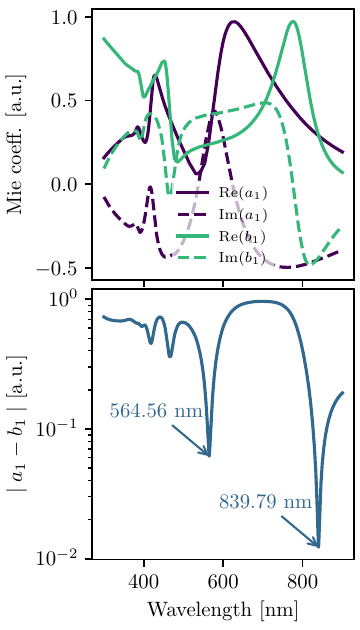}};
            \node[anchor=north west, xshift=3pt, yshift=-3pt] at (img.north west) {\textbf{(a)}};
        \end{tikzpicture}
    \end{minipage}
    \hfill
    \begin{minipage}[t]{0.28\textwidth}
        \vspace{0pt} 
        \centering
        \begin{tikzpicture}[inner sep=0]
            \node[anchor=south west] (img) at (0,0) {\includegraphics[width=\textwidth]{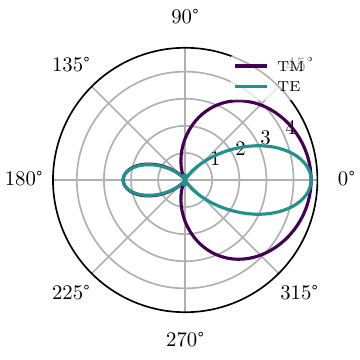}};
            \node[anchor=north west, xshift=3pt, yshift=-3pt] at (img.north west) {\textbf{(b)}};
        \end{tikzpicture}
    
        \begin{tikzpicture}[inner sep=0]
            \node[anchor=south west] (img) at (0,0) {\includegraphics[width=\textwidth]{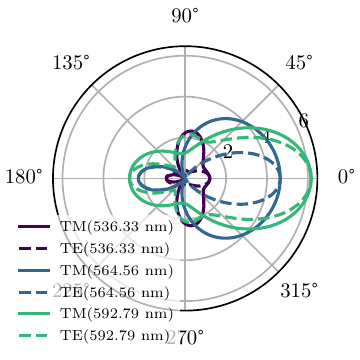}};
            \node[anchor=north west, xshift=3pt, yshift=-3pt] at (img.north west) {\textbf{(d)}};
        \end{tikzpicture}
    \end{minipage}
    \hfill
    \begin{minipage}[t]{0.28\textwidth}
        \vspace{0pt} 
        \centering
        \begin{tikzpicture}[inner sep=0]
            \node[anchor=south west] (img) at (0,0) {\includegraphics[width=\textwidth]{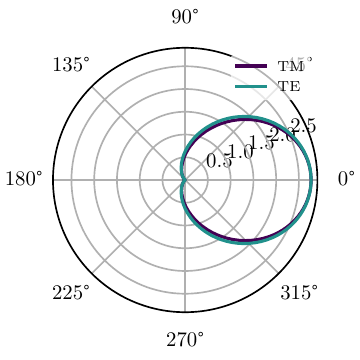}};
            \node[anchor=north west, xshift=3pt, yshift=-3pt] at (img.north west) {\textbf{(c)}};
        \end{tikzpicture}
    
        \begin{tikzpicture}[inner sep=0]
            \node[anchor=south west] (img) at (0,0) {\includegraphics[width=\textwidth]{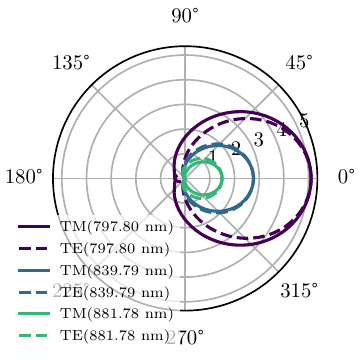}};
            \node[anchor=north west, xshift=3pt, yshift=-3pt] at (img.north west) {\textbf{(e)}};
        \end{tikzpicture}
    \end{minipage}
    \hfill\hfill
    \caption{(a) Real and imaginary parts of (top) and absolute difference $\mid a_1-b_1\mid $ between (bottom) the $a_1$ and $b_1$ Mie coefficients for a Si nanoparticle of 100\,nm radius surrounded by a medium with $n=1.333$ (such as water) within the visible range (300 to 900\,nm). $\mid a_1-b_1\mid$ is locally smallest at 564.56\,nm and 839.79\,nm, and thus these are the wavelengths where the Kerker condition $a_1=b_1$ is most closely fulfilled. The TM and TE components of the intensity scattered by the same Si nanoparticle for such incident wavelengths where $a_1 \approx b_1$ are shown in (b) and (c), respectively. In (c), a pure zero-backward scattering behavior is observed. On the other hand, in (b), only the TM component shows zero-backward scattering. (d) and (e) present a sensitivity analysis of panels (b) and (c), respectively, showing the TM and TE components for wavelengths that are $\pm$5\% around the nominal value. These results confirm that the wavelengths closest to satisfying $a_1=b_1$ provide the optimal balance of strong forward scattering with minimal backward scattering.}
    \label{fig:part-2-intensity}
\end{figure*}

The simultaneous excitation of electric and magnetic modes in high-index nanoparticles enables their interaction and, crucially, their interference. This interplay leads to exotic phenomena such as directional scattering (see \textbf{Figure \ref{fig:part-2-intensity}}). This phenomenon, which emerges when the Kerker condition $a_1 = b_1$ is satisfied \cite{kerker1983electromagnetic, fu2013directional, Liu:18}, permits enhanced directional power concentration advantageous for sensing applications. Such behavior is unattainable in the small metallic nanoparticles studied previously, as their optical response is restricted to a single electric dipolar mode.

\section*{Conclusions}

In this work, we have performed a comparative analysis of the optical resonance characteristics of Gold and high-index semiconductor nanoparticles to evaluate their viability for refractive index sensing.

Simulations of Au nanoparticles revealed a strong dependence of the LSPR on the refractive index of the surrounding medium, yielding a sensitivity of $\approx$150\,nm/RIU. However, the non-trivial dependence of the spectral shift on the refractive index and the inherent ohmic losses of metals impose limitations on precision. Furthermore, the acute sensitivity of the resonance position to geometric variations implies that stringent monodispersity is required to prevent size distribution from obscuring environmental refractive index signals.

In contrast, our investigation of high-index semiconductor nanoparticles (e.g., Si, Ge, GaAs) demonstrated that these dielectric structures support both electric and magnetic Mie resonances while maintaining negligible absorption losses. The detailed study of the extinction efficiency in Silicon nanoparticles confirmed the coexistence of electric and magnetic dipoles and quadrupoles. Crucially, the interference between the dipolar electric and magnetic modes ($a_1$ and $b_1$) facilitates directional scattering phenomena when $a_1\approx b_1$, a feature fundamentally unattainable in small metallic nanoparticles.

Ultimately, while Au nanoparticles offer established bio-compatibility and high sensitivity, high-index dielectrics provide a lower-loss alternative with richer spectral features. The ability to engineer directional scattering in semiconductors opens new avenues for sensing modalities that rely on signal directionality rather than pure spectral shifting, potentially enhancing the signal-to-noise ratio in complex sensing environments.

\section*{Declaration of Competing Interest}

The author declares no competing interests.

\section*{Acknowledgments}

The author thanks Professor Braulio García Cámara for providing a basic scattering simulation code and for helpful discussions.

\section*{Data and Code availability}

The data and code used to generate the results in this study is available on \textit{GitHub} at \url{https://github.com/bfrangi/nanoparticle-scattering}.
An archived, citable version of the code corresponding to this publication is available on \textit{Zenodo} (\href{https://doi.org/10.5281/zenodo.18069314}{10.5281/zenodo.18069314}).

\bibliography{sample}

@article{willets2007localized,
    author = {Katherine A. Willets and Richard P. Van Duyne},
    title = "Localized Surface Plasmon Resonance Spectroscopy and Sensing", 
    journal= "Annual Review of Physical Chemistry",
    year = "2007",
    volume = "58",
    pages = "267-297",
    doi = "10.1146/annurev.physchem.58.032806.104607",
    publisher = "Annual Reviews",
    issn = "1545-1593",
}

@article{kuznetsov2016optically,
    author = {Arseniy I. Kuznetsov  and Andrey E. Miroshnichenko  and Mark L. Brongersma  and Yuri S. Kivshar  and Boris Luk’yanchuk },
    title = {Optically resonant dielectric nanostructures},
    journal = {Science},
    volume = {354},
    number = {6314},
    year = {2016},
    doi = {10.1126/science.aag2472},
}

@article{saha2012gold,
    title={Gold nanoparticles in chemical and biological sensing},
    author={Saha, Krishnendu and Agasti, Sarit S and Kim, Chaekyu and Li, Xiaoning and Rotello, Vincent M},
    journal={Chemical Reviews},
    volume={112},
    number={5},
    pages={2739--2779},
    year={2012},
    publisher={ACS Publications},
    doi={10.1021/cr2001178}
}

@article{kerker1983electromagnetic,
    author = {M. Kerker and D.-S. Wang and C. L. Giles},
    journal = {J. Opt. Soc. Am.},
    number = {6},
    pages = {765--767},
    publisher = {Optica Publishing Group},
    title = {Electromagnetic scattering by magnetic spheres},
    volume = {73},
    month = {Jun},
    year = {1983},
    doi = {10.1364/JOSA.73.000765},
}

@article{fu2013directional,
    title={Directional visible light scattering by silicon nanoparticles},
    author={Fu, Yuan Hsing and Kuznetsov, Arseniy I and Miroshnichenko, Andrey E and Yu, Ye Feng and Luk’yanchuk, Boris},
    journal={Nature Communications},
    volume={4},
    number={1},
    pages={1527},
    year={2013},
    publisher={Nature Publishing Group},
    doi={10.1038/ncomms2538}
}

@misc{bernat_frangi_2025_18069231,
    author       = {Bernat Frangi},
    title        = {Mie Scattering Simulator},
    month        = {Dec},
    year         = {2025},
    howpublished = {Zenodo},
    doi          = {10.5281/zenodo.18069231},
}

@article{Boriskina:17,
    author = {Svetlana V. Boriskina and Thomas Alan Cooper and Lingping Zeng and George Ni and Jonathan K. Tong and Yoichiro Tsurimaki and Yi Huang and Laureen Meroueh and Gerald Mahan and Gang Chen},
    journal = {Adv. Opt. Photon.},
    number = {4},
    pages = {775--827},
    publisher = {Optica Publishing Group},
    title = {Losses in plasmonics: from mitigating energy dissipation to embracing loss-enabled functionalities},
    volume = {9},
    month = {Dec},
    year = {2017},
    doi = {10.1364/AOP.9.000775},
}

@Article{Khurgin2015,
    author={Khurgin, Jacob B.},
    title={How to deal with the loss in plasmonics and metamaterials},
    journal={Nature Nanotechnology},
    year={2015},
    month={Jan},
    day={01},
    volume={10},
    number={1},
    pages={2--6},
    issn={1748-3395},
    doi={10.1038/nnano.2014.310},
}

@article{Kivshar:17,
    author = {Yuri Kivshar and Andrey Miroshnichenko},
    journal = {Opt. Photon. News},
    number = {1},
    pages = {24--31},
    publisher = {Optica Publishing Group},
    title = {Meta-Optics with Mie Resonances},
    volume = {28},
    month = {Jan},
    year = {2017},
    doi = {10.1364/OPN.28.1.000024},
}

@article{doi:10.1021/cr100313v,
    author = {Mayer, Kathryn M. and Hafner, Jason H.},
    title = {Localized Surface Plasmon Resonance Sensors},
    journal = {Chemical Reviews},
    volume = {111},
    number = {6},
    pages = {3828-3857},
    year = {2011},
    doi = {10.1021/cr100313v},
    note = {PMID: 21648956},
}

@article{Liu:18,
    author = {Wei Liu and Yuri S. Kivshar},
    journal = {Opt. Express},
    number = {10},
    pages = {13085--13105},
    publisher = {Optica Publishing Group},
    title = {Generalized Kerker effects in nanophotonics and meta-optics},
    volume = {26},
    month = {May},
    year = {2018},
    doi = {10.1364/OE.26.013085},
}

@Article{C1CS15237H,
    author = {Dreaden, Erik C. and Alkilany, Alaaldin M. and Huang, Xiaohua and Murphy, Catherine J. and El-Sayed, Mostafa A.},
    title  = {The golden age: gold nanoparticles for biomedicine},
    journal  = {Chem. Soc. Rev.},
    year  = {2012},
    volume  = {41},
    issue  = {7},
    pages  ={2740--2779},
    publisher  = {{T}he {R}oyal {S}ociety of {C}hemistry},
    doi  = {10.1039/C1CS15237H},
}

\end{multicols}

\end{document}